\documentclass[aps,prl,twocolumn,superscriptaddress,amsmath,amssymb,aps]{revtex4-2}

\usepackage{graphicx}
\usepackage[normalem]{ulem} 
\usepackage{xcolor}  		
\usepackage{dcolumn}
\usepackage{bm}

\begin{document}

\title{Burgers turbulence in the Fermi-Pasta-Ulam-Tsingou chain}

\author{Matteo Gallone}
\email{matteo.gallone@sissa.it}
\affiliation{SISSA, Via Bonomea 265, 34136 Trieste, Italy}
%
\author{Matteo Marian}%
 \email{matteo.marian@studenti.units.it}
\affiliation{ 
Department of Physics, University of Trieste, Via A. Valerio 2, 34127 Trieste, Italy} 

\author{Antonio Ponno}
 \email{ponno@math.unipd.it}
\affiliation{%
Department of Mathematics ``T. Levi-Civita'', University of Padova, Via Trieste 63, 35121 Padova, Italy}

\author{Stefano Ruffo}%
 \email{ruffo@sissa.it}
\affiliation{SISSA, Via Bonomea 265, 34136 Trieste, Italy}
\address{INFN Sezione di Trieste}
\address{
ISC-CNR, via Madonna del Piano 10, 50019 Sesto Fiorentino (Firenze)
}
%

\date{\today}

\begin{abstract}
We prove analytically and show numerically that the dynamics of the Fermi-Pasta-Ulam-Tsingou chain is characterised by a transient Burgers turbulence regime on a wide range of time and energy scales. This regime is present at long wavelengths and energy per particle small enough that equipartition is not reached on a fast time scale. In this range, we prove that the driving mechanism to thermalisation is the formation of a shock  that can be predicted using a pair of generalised Burgers equations. We perform a perturbative calculation at small energy per particle, proving that the energy spectrum of the chain $E_k$ decays as a power law, $E_k\sim k^{-\zeta(t)}$, on an extensive range of wavenumbers $k$. We predict that $\zeta(t)$ takes first the value $8/3$ at the Burgers shock time, and then reaches a value close to $2$ within two shock times. The value of the exponent $\zeta=2$ persists for several shock times before the system eventually relaxes to equipartition. During this wide time-window, an exponential cut-off in the spectrum is observed at large $k$, in agreement with previous results. Such a scenario turns out to be universal, i.e. independent of the parameters characterising the system and of the initial condition, once time is measured in units of the shock time. 
\end{abstract}

\maketitle

\emph{-- Introduction.} Understanding the route to thermalisation of an isolated physical system is a fundamental problem in statistical mechanics. The behaviour close to equilibrium has been widely understood, while the situation is much more complex when the system is initialised far from equilibrium~\cite{Balescu97}. Historically, the first system that did not display thermalisation on the 
observation time scale was the Fermi-Pasta-Ulam-Tsingou (FPUT) chain~\cite{FPU55,GallavottiBook}. The authors studied, in a computer simulation, a  simple one-dimensional model of nonlinearly interacting classical particles with the aim of observing the rate of thermalisation. Instead of the expected trend to equilibrium, they observed a ``recurrent'', quasi-periodic behaviour and a lack of energy equipartition among the Fourier modes. An interpretation of such a ``FPUT paradox'' in terms of Korteweg-de Vries (KdV) solitons was provided in~\cite{ZK65}.
A complementary interpretation, based on the so-called KAM theory~\cite{KAM}, was proposed in~\cite{IC66}, where the FPUT phenomenon was linked to the criterion of ``resonance overlap'' for the transition to chaos. The problem of thermalisation is still a subject of active investigation: phenomena related to the FPUT recurrence have been observed in several systems, from graphene resonators~\cite{Midt} to nonlinear phononic~\cite{Cao} and photonic \cite{Piera} systems, from trapped cold atoms~\cite{Kino} to Bose-Einstein condensates~\cite{Villa,Dan}.

The FPUT model consists of $N$ unit masses sitting on a one-dimensional lattice and connected by nearest-neighbour non-linear springs. The Hamiltonian of the $\alpha$+$\beta$ FPUT model is 
\begin{equation}
\label{eq:H}
H=\sum_{j=1}^N \left[\frac{p_j^2}{2}+V(q_{j+1}-q_j)\right]\ ,
\end{equation}
where $V(z)=\frac{z^2}{2}+\alpha\frac{z^3}{3}+\beta\frac{z^4}{4}$, 
$q_j$ is the displacement from equilibrium of the $j$-th mass and $p_j$ its momentum. 

If the non-linear part of the interaction vanishes, i.e. $\alpha=\beta=0$, the dynamics of the \emph{Fourier Energy Spectrum (FES)} becomes trivial, since no exchange of energy among the Fourier modes is possible. Thermalisation is driven by nonlinearity, which couples the modes causing energy exchange.
However, mode-coupling takes place also in nonlinear integrable systems, such as the Toda chain~\cite{HenonToda}, where no thermalisation occurs. The approach to equilibrium of integrable systems has been recently studied in~\cite{Goldfriend}. 

A generic feature of both integrable and quasi-integrable one-dimensional systems is the presence of an exponentially decaying \emph{FES}~\cite{Parisi82,Vulpiani83}. Moreover, for the FPUT model, the long wavelength modes form a ``packet'' of size $\varepsilon^{1/4}$~\cite{Shep,Po03}, where $\varepsilon$ is the specific energy $\varepsilon=E/N$. This scenario describes the behavior of the \emph{FES} of quasi-integrable systems on time scales increasing as inverse power-laws of $\varepsilon$~\cite{Cohen,BP11}, whereas for integrable systems the \emph{FES} remains exponentially localised for all times. 
It is known \cite{IC66} that the FPUT chain relaxes to equipartition on a faster time scale at sufficiently large specific energies \cite{Vulpiani85,Ooms}. More recently, it was shown that relaxation takes place also at smaller energies, see Ref. \cite{BP11} for a discussion. Relaxation eventually occurs also in the energy range studied in this Letter.

\begin{figure}[h!]
\includegraphics[width=\columnwidth]{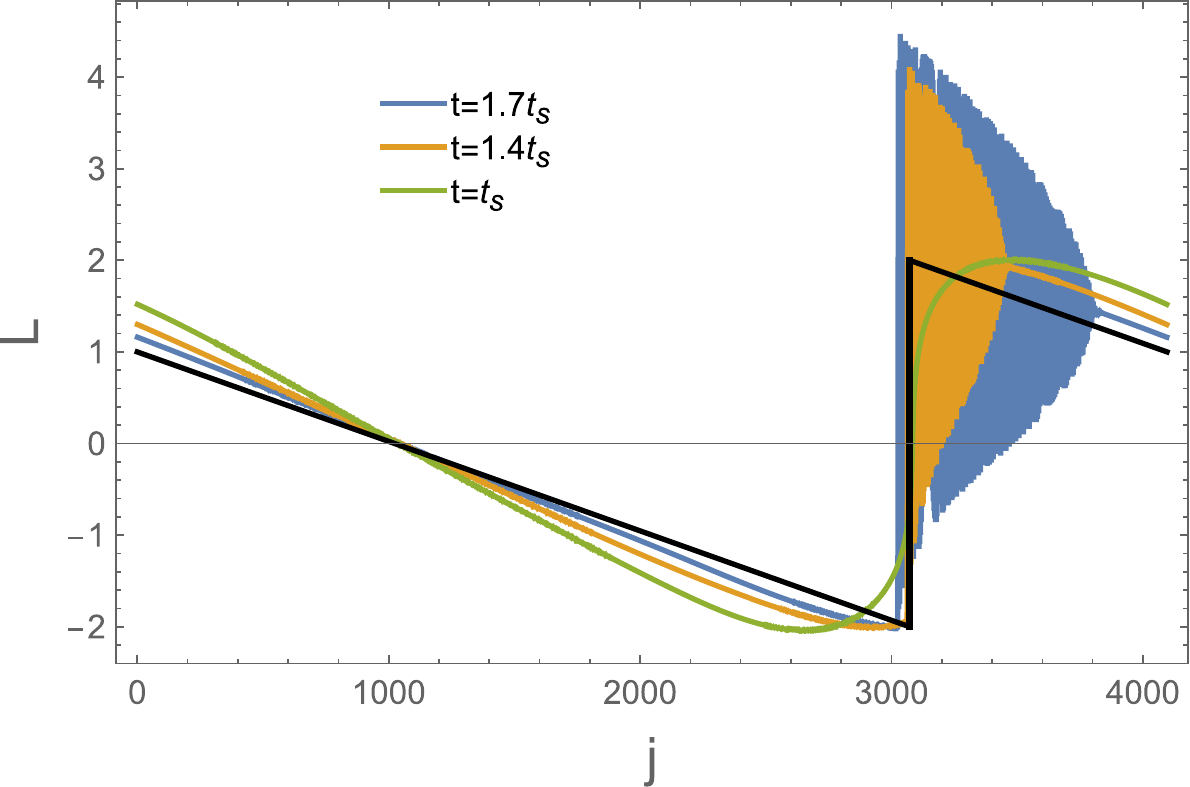}
\caption{Numerical simulation of the FPUT model (\ref{eq:H}) for $\varepsilon=0.07$, $N=4096$, $\alpha=1$, $\beta=1/2$. The coloured solid lines are the profiles corresponding to a left Travelling Wave Excitation (TWE) plotted at the shock time 
$t_s$, formula (\ref{eq:ts}), and at later times. Notice the evolution towards a sawtooth profile (black solid line) followed by fast oscillations (discussed in the text).
}
\label{fig:sawtooth}
\end{figure}

In this Letter, we study the FPUT chain in a regime where the specific energy $\varepsilon$ is large enough that mode-coupling acts on a wide range of long wavelength modes, but is still small enough to slow down thermalisation. In this regime the long wavelength \emph{FES} turns out to be a scale invariant power-law, which motivates 
the use of the term ``turbulence'' to describe this phenomenon. The range of involved modes is of the size of the ``packet'' quoted above. Our analysis begins with the observation that, in this regime, the time evolution of an initial wave leads to the formation of a ``shock'', as shown in Fig.~\ref{fig:sawtooth}. This behaviour was first described in~\cite{PRK95} and is strongly related to the non dispersive limit 
of the KdV equation~\cite{ZK65,Lax80}, i.e. the inviscid Burgers equation. In this Letter, we show that the dynamics of the FPUT chain, in a specific time range, is well described by a pair of \emph{generalised} Burgers equations.

Our approach allows us to derive rigorously and compute analytically some properties of the \emph{FES} in a wide range of specific energy values.

\begin{figure}[h!]
\includegraphics[width=\columnwidth]{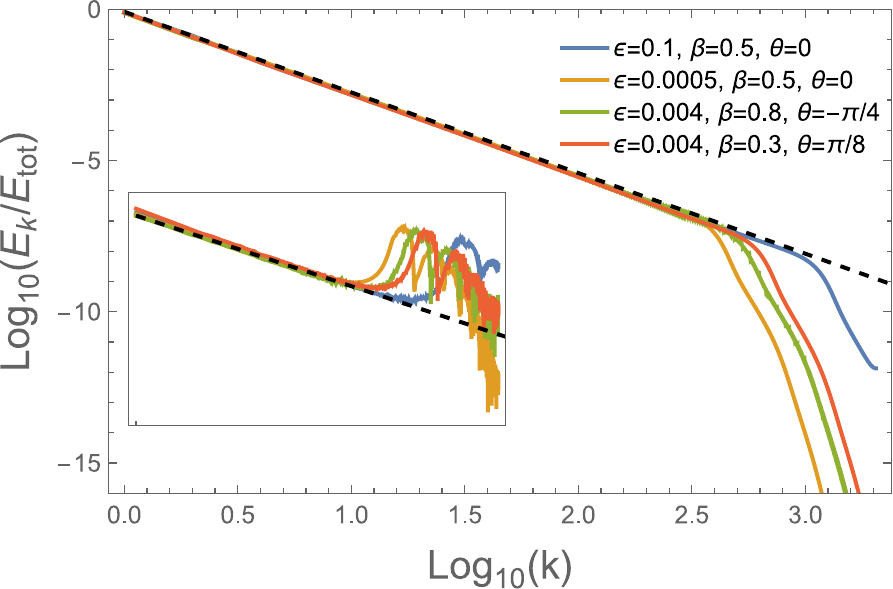}
\caption{Normalised \emph{FES} of the FPUT model (\ref{eq:H}) for $\alpha=1$, different values of $\beta$ and $N=4096$ at the shock time $t_s$, formula (\ref{eq:ts}). The initial condition is (\ref{eq:indat}) with different values of $\theta=\varphi-\pi/4$ and 
$\varepsilon$. The dashed line is the theoretical prediction (\ref{eq:fesgen2}), $E_k/E\simeq 0.8 k^{-8/3}$. 
Notice the exponential cut-off at large $k$.
Inset: \emph{FES} at $4 t_s$ for the same initial conditions. The dashed line is the theoretical prediction $E_k \sim k^{-2}$.}
\label{fig:Spectra}
\end{figure}

\begin{figure}[h!]
\includegraphics[width=\columnwidth]{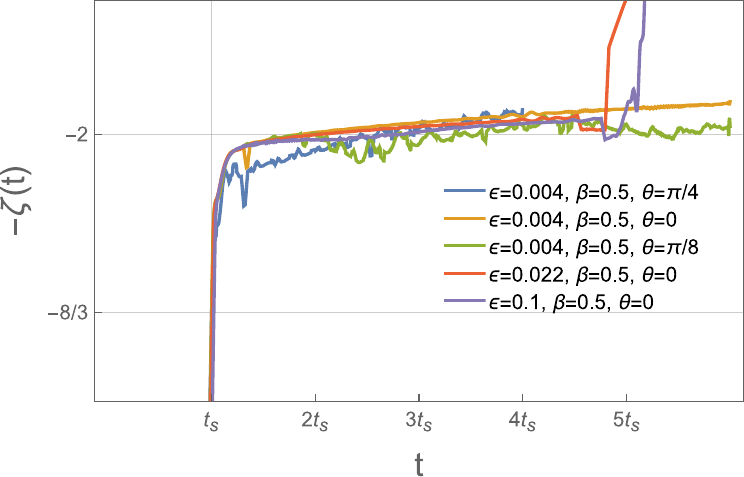}
\caption{Slope $-\zeta(t)$ of the power-law that interpolates the \emph{FES} at small $k$ and for $N=4096$, see formula (\ref{eq:ZetaExponent}).  One should remark that the data collapse follows from measuring the time in units of $t_s$ (\ref{eq:ts}), which incorporates all the different values of the initial conditions and the parameters of the Hamiltonian. } 
\label{fig:CoeffAng}
\end{figure}

\noindent
\emph{-- Main results.}  Corresponding to an initial excitation of the longest wavelength, we determine a window of low modes where the \emph{FES} scales with an inverse time-dependent power-law
\begin{equation}\label{eq:ZetaExponent}
	E_k \sim k^{-\zeta(t)}\ \ ;\ \ k_0\leq k\leq k_c\ ,
\end{equation}
with $k_0$ and $k_c$ slowly depending on time. The window $[k_0,k_c]$ scales with the number of particles $N$, i.e. is \emph{extensive} in $N$,
and $k_0$ is of order $1$. We find a shock time-scale $t_s$  that characterises a fast energy transfer from the initially excited mode $k=1$ to the higher ones. The value of the exponent $\zeta(t)$ at $t_s$ is $\zeta(t_s)=8/3$, as 
shown in Fig.~\ref{fig:Spectra}. We determine analytically both $t_s$ and the corresponding value of the exponent in terms of the underlying Burgers dynamics of the system. We then observe that within a time $\sim 2t_s$, the exponent $\zeta(t)$ decreases to a value of about $2$, see Fig.~\ref{fig:CoeffAng} and the inset of Fig.~\ref{fig:Spectra}.  
The \emph{FES} $E_k\sim k^{-2}$ is preserved up to four shock times, after which the power law structure is lost and the system eventually reaches the statistical equilibrium characterised by an almost flat \emph{FES} (energy equipartition), as shown
in Fig.~\ref{fig:CoeffAng} by the growth of the slope at later times. The whole phenomenology observed resembles the one of turbulence in fluids~\cite{FP}, with an inertial range $[k_0,k_c]$ over which the \emph{FES} displays a power law decay. However, in absence of energy injection and dissipation, we are here in presence of a transient turbulence phenomenon. Moreover, it must be stressed that the values of the exponent $8/3$ at $t_s$ and $2$ at later times, are clear signatures of an evolution guided by the \emph{integrable} Burgers dynamics~\cite{FF}. Finally, like in fluid turbulence, we observe an exponential decay of the~\emph{FES} beyond the inertial range, i.e. for  values of $k>k_c$. In fluids this is due to a small scale balance between nonlinearity and dissipation~\cite{FP}, whereas in our case the role of dissipation is played by dispersion. In addition, as for decaying turbulence in fluids, after the transient turbulence regime we observe that the exponential fall-off disappears and the \emph{FES} becomes flat, eventually leading to energy equipartition.
The phenomenology treated here does not fall into the range of applicability of the so-called (weak) wave 
turbulence~\cite{Onorato,Lvov2018}, which would require an unfitting assumption of weak nonlinearity.

\emph{-- Model, initial conditions and continuum approximation.} 
All the details of the following analytical derivation are reported in the Supplemental Material~\cite{suppmat} (see~\cite{hptpde} for the mathematical framework). 

For the FPUT model \eqref{eq:H} we choose periodic boundary conditions: $q_N=q_0$ and $p_N=p_0$. 
Defining the Fourier coefficient $\hat{q}_k \;=\; \frac{1}{\sqrt{N}}\sum_{j=1}^N q_j e^{\imath 2 \pi k j/N}$ of the displacements $q_j$,
and similarly for the momenta $p_j$, the energy of the linearised system is consequently written as
\begin{equation}
		H_{lin}=\sum_{k=1}^N E_k \, ,  \qquad\, E_k:=\frac{|\hat{p}_k|^2}{2}+\frac{\omega_k^2 |\hat{q}_k|^2}{2} 
\end{equation}
where $\omega_k=2\sin(\pi k/N)$ and $E_k$ is the energy of mode $k$.

We consider the two-parameter family of initial data
\begin{subequations}
\label{eq:indat}
\begin{eqnarray}
	q_j(0)&=& A\cos\varphi\sin\left(\frac{2 \pi j}{N} \right)\ ;\\
	p_j(0)&=& \omega_1A \sin\varphi\cos\left(\frac{2 \pi j}{N}\right)\ ,	
\end{eqnarray}
\end{subequations}
for $j=1,\dots,N$. Here, $A>0$ and $0\leq\varphi\leq\pi/2$ are the amplitude and the phase of the initial excitation. Varying the phase from 
$\varphi=0$ to $\varphi=\pi/2$, we tune the kinetic energy of the initial condition (\ref{eq:indat}).
The value $\varphi=\pi/4$ corresponds to a left Travelling Wave Excitation (TWE), around which we explore a large neighborhood. The specific energy $\varepsilon=E/N$ can be written in terms of $A$ and $\varphi$, for large $N$, as
\begin{equation}
\label{eq:epsa}
\varepsilon=a^2+\frac{3\beta}{2}\left(a\cos\varphi\right)^4\ ,\ \ a=\frac{\pi A}{N}\ .
\end{equation}
In order to study the evolution of the initial condition (\ref{eq:indat}) in the continuum limit $N\to\infty$, at fixed small
$a$, we first introduce two fields $Q(x,\tau)$ and $P(x,\tau)$ of spatial period one, such that
$q_j(t)=NQ(j/N,t/N)$, $p_j(t)=P(j/N,t/N)$. 

In order to separate the right from the left motion at zero order in the small parameter $a$, we then introduce the ``left'' and ``right'' fields $L=(Q_x+P)/(a\sqrt{2})$, $R=(Q_x-P)/(a\sqrt{2})$, where partial derivatives are denoted by subscripts. 

The evolution equations in the continuum limit read  $L_\tau=L_x+O(a)$, $R_\tau=-R_x+O(a)$, which in the harmonic limit $a\to0$ uncouple into the left and right translations of the initial conditions $L_0(x)$ and $R_0(x)$. It follows from (\ref{eq:indat}) that $L_0$ has maximal amplitude for $\varphi=\pi/4$, when $R_0=0$, which defines the left TWE. The equations of motion display the symmetry $\varphi\to-\varphi$, $L\to R$.

Since the equations for $L$ and $R$ are 
nonlinearly coupled for any $a>0$, we 
build up a transformation $\mathcal{C}_a:(L,R)\mapsto(\lambda,\rho)$ of the fields matching the identity
for $a\to0$ and such that the evolution equations of the new fields $\lambda$ and $\rho$ turn out to be decoupled to
order $a^2$ included. A rather long computation yields \cite{suppmat}
\begin{subequations}
\label{eq:larosys}
\begin{equation}
\label{eq:laroeq}
\lambda_\tau=\Phi(\lambda)\lambda_x\ \ ;\ \ 
\rho_\tau=-\Phi(\rho)\rho_x\ ;
\end{equation}
\begin{equation}
\label{eq:Phi}
\Phi(\lambda)=\frac{a\alpha}{\sqrt{2}}\lambda+
\frac{3a^2\alpha^2}{4}\left(\frac{\beta}{\alpha^2}-\frac{1}{2}\right)\lambda^2\ ,
\end{equation}
\end{subequations}
with initial conditions $(\lambda_0,\rho_0)=\mathcal{C}_a(L_0,R_0)$.

Due to the form of the nonlinearity, equations \eqref{eq:larosys} reduce to a pair of Burgers equations if $\beta=\frac{\alpha^2}{2}$ or, otherwise, to a pair of generalised  Burgers equations.

\emph{-- Shock time and universal FES.}
The equations of motion (\ref{eq:laroeq}) for the left and right fields $\lambda(x,\tau)$ and $\rho(x,\tau)$ have the form of two  uncoupled inviscid, generalised Burgers equations. 
Their solution exists in a finite time interval $[0,\tau_s[$, where $\tau_s$ is the shock time~\cite{suppmat}.

Taking into account the time rescaling $t_s=N\tau_s$, we obtain the following expression for the FPUT shock time $t_s$
\begin{equation}
\label{eq:ts}
t_s=\left(\frac{N}{2\pi\sqrt{2}a\alpha}\right)\frac{F(\mu)}{\cos\theta}\ ,
\end{equation}
where the function $F(\mu)$ and the auxiliary parameter $\mu$ are given by 
\begin{equation}
\label{eq:Fmu}
F(\mu)=\sqrt{\frac{32\mu^2}{\sqrt{1+32\mu^2}-1+16\mu^2}}\frac{4}{\sqrt{1+32\mu^2}+3}\ ;
\end{equation}
\begin{equation}
\label{eq:mu}
\mu=\frac{a\alpha}{2\sqrt{2}}\cos\theta\left[
\tan^2\theta-4\tan\theta+6\left(\frac{\beta}{\alpha^2}-\frac{1}{2}\right)\right]\, .
\end{equation}
Formula~(\ref{eq:ts}) is valid for $a$ small enough and 
$-\pi/4\leq \theta\leq\pi/4$, where $\theta=\varphi-\pi/4$. 

In order to estimate the \emph{FES} of the FPUT model at the shock time $t_s$, we generalise the procedure of~\cite{FF} and compute the exact solution of \eqref{eq:laroeq} in Fourier space
\begin{equation}
\label{eq:uk}
\hat \lambda_k(\tau)=\frac{1}{\imath 2\pi k}\oint \lambda_0'(x)e^{-\imath2\pi k[x-\tau \Phi(\lambda_0(x))]}dx\ ,
\end{equation}
and the analogous one for $\hat{\rho}_k(\tau)$.
Then, for a general class of initial conditions the method of (degenerate) stationary phase applied to
the integral (\ref{eq:uk}) yields $|\hat \lambda_k(\tau_s)|^2\sim C\ k^{-8/3}$ 

for large $k$, where $C$ is an explicit constant independent of $k$. It also turns out that 
$|\hat\rho_k(\tau_s)|^2$ is smaller than $|\hat \lambda_k(\tau_s)|^2$, the smaller the closer $\theta$ is to $\pi/4$, equality holding for $\theta=\pm\pi/4$. Taking into account the relation 
\begin{equation}
	E_k(t_s)\propto |\hat \lambda_k(\tau_s)|^2+|\hat\rho_k(\tau_s)|^2 \, ,
\end{equation}
 we derive the normalised \emph{FES} of the FPUT system
as 
\begin{equation}
\label{eq:fesgen2}
\frac{E_k(t_s)}{\sum_kE_k(t_s)}=(0.7787\dots)\ k^{-8/3}\ .
\end{equation}
Notice that the shock time $t_s$ incorporates all the dependencies of the \emph{FES} on the parameters of the system and of the initial conditions, so that the spectrum (\ref{eq:fesgen2}) is indeed \emph{universal}.

\begin{figure}[h!]
\includegraphics[width=\columnwidth]{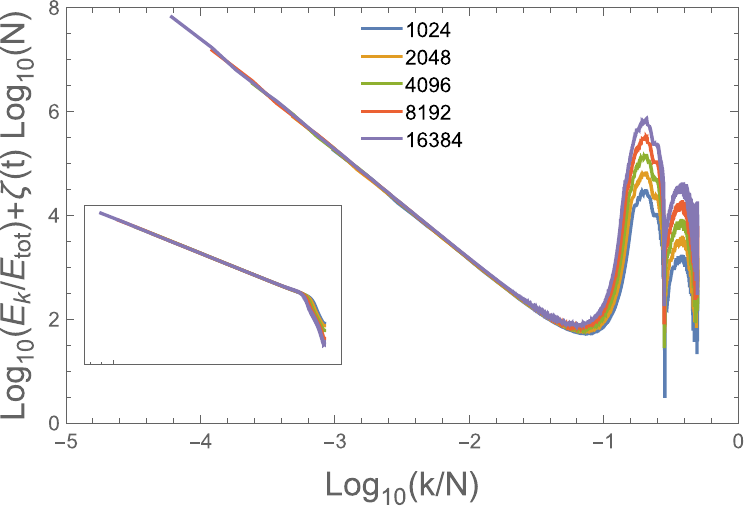}
\caption{\emph{FES} vs. $k/N$, left TWE at $4 t_s$, $\alpha=1$, $\beta=1/2$, $\varepsilon= 0.05$ and different values of $N$. Inset: the same at $t_s$.}
\label{fig:fesspec}
\end{figure}

We have performed massive numerical simulations~\cite{numerics} of the FPUT system (\ref{eq:H})-(\ref{eq:indat}). The \emph{FES} at the shock time (\ref{eq:ts}) is displayed in Fig.~\ref{fig:Spectra}, for different initial conditions. The universal \emph{FES} (\ref{eq:fesgen2}) works over $6$ to $7$ orders of magnitude in mode energy, and the scenario is robust over three  orders of magnitudes in specific energy. Fig.~\ref{fig:Spectra} also shows the presence of an exponential cut-off beyond $k_c$, consistently with the theory of~\cite{Parisi82}. We have verified that $k_c/N\propto\varepsilon^{1/4}$, in agreement with~\cite{Shep,BGG04}, so that the scaling region in $k$ is extensive, as shown in Fig.\ref{fig:fesspec}.

\emph{-- Beyond the shock time.}
The solution of the generalised Burgers equations (\ref{eq:larosys}) no longer exists for times $\tau>\tau_s=t_s/N$, due to
a local divergence of the derivatives of the fields. Such a ``gradient catastrophe'' implies a transfer of energy to the highest Fourier modes of wavelength $\sim1/N$, so that a global continuum limit no longer holds after the shock. For a correct continuum description of the shock region, higher order derivatives
of the fields must be taken into account, which replaces the Burgers equations with a pair of KdV equations~\cite{hptpde,PB05,BP06,GPR}. However, far from the shock region, the Burgers equation still describes the FPUT dynamics.
Indeed, let us consider the left TWE $\lambda(x,0)=2\cos(2\pi x)$ with $\beta=1/2$ in order to eliminate the quadratic term in $a$ in Eq.~(\ref{eq:Phi}). In this case system (\ref{eq:larosys}) yields the Burgers equation
$\lambda_\tau=(a\alpha/\sqrt{2})\lambda\lambda_x$, whose solution is obtained from the implicit equation $\lambda=2\cos\left(2\pi(x+(a\alpha/\sqrt{2})\lambda \tau)\right)$.
The initial cosine is progressively deformed into a sawtooth profile $\sigma(x)$ with the discontinuity at $x=3/4$ (the point in which the
initial cosine vanishes and the profile has positive derivative) and slope $-4$. Performing a Fourier transform one finds that
\begin{equation} 
\label{eq:saw}
\sigma(x)=\sum_{k\neq0}\frac{2}{\imath\pi k}e^{\imath2\pi k(x+1/4)}\ .
\end{equation}
It can be shown that the time needed for the position of the maximum of the initial cosine to reach the node at
$x=3/4$ is $(\pi/2)\tau_s$, thus larger than the shock time $\tau_s$.  At the shock time $\tau_s$ the spatial derivative of $\lambda$ becomes infinite in the Burgers equation, huge but finite on the lattice due to dispersion. The formation of the sawtooth profile then follows in time
the creation of the shock. Heuristically, after this formation, one can decompose the wave profile as $\lambda(x,\tau)=\sigma(x)+r(x,\tau)$, where the deviation $r$ with respect to the sawtooth profile (\ref{eq:saw}) is smooth. The Fourier coefficients of $\sigma(x)$ decay 
as $1/k$, while those of the smooth deviation $r$ can be shown to decay faster~\cite{cinfinity}. Therefore, the \emph{FES} of 
$\lambda$ is dominated by $|\hat\sigma_k|^2\propto k^{-2}$.  This heuristic argument can be verified in numerical experiments by measuring
the slope of the \emph{FES} after the shock time. The time evolution of the slope is shown in Fig.~\ref{fig:CoeffAng}: one observes an
extended time domain (approximately from two to four shock times) where the slope remains close to $-2$. The relevance of the scaling exponent
$2$ for Burgers turbulence was already established in~\cite{FF} and further analyzed in~\cite{Kida}. Although the numerical determination of the slope for later times becomes much harder, it can be seen that it eventually increases, detecting a trend to equipartition, which corresponds to a vanishing slope and the disappearance of the exponential fall off. It is also important to highlight the data ``collapse'', which is a consequence of measuring time in
units of the shock time $t_s$ (\ref{eq:ts}). In the inset of Fig.~\ref{fig:Spectra} we display the \emph{FES} at $4 t_s$ in order to confirm that $\zeta=2$. We observe the additional presence of a peak at large $k$. We plot in log-log scale the energy spectrum versus $k$ adjusting a line with slope $-2$ on the experimental data.

In a statistical mechanical perspective, the \emph{FES} vs. $k/N$ is reported in Fig.~\ref{fig:fesspec}. The proportionality to $N$ of the power-law window is evident, which implies that Burgers turbulence is a relevant phenomenon in the thermodynamic limit
of the FPUT system.

In order to explain the presence of the peaks in the \emph{FES} of Fig.~\ref{fig:Spectra}, we go back to the analysis of Fig.~\ref{fig:sawtooth}. We display there the numerical profiles of the left TWE, i.e.
$(q_{j+1}-q_j+p_j)/(\sqrt{2}a)$ vs. $j$, up to a suitable Galileian translation \cite{suppmat},
for three different times. 
We clearly observe the formation of the sawtooth profile, and the fast oscillations near the discontinuity of the profile. These
oscillations have been studied by various authors~\cite{Lax80} in the context of the non dispersive limit of the KdV
equation. In our approach, this oscillatory part of the profile is included in the smooth deviation $r$ from the sawtooth $\sigma$. We think that these
oscillations are the main feature of the spatial profile which determines the observed peak in the \emph{FES} at large $k$.

 Short-wavelength oscillations are found also in the Galerkin-truncated Burgers equation. These oscillations, called ``tygers'' (see e.g. \cite{Ray}), are however of different nature with respect to the ones observed in FPUT: ``tygers'' are due to the Galerkin truncation, while the ones we observe in the FPUT are due to the small dispersion term of the approximating KdV dynamics. 
Nevertheless, phenomena similar to the ones that give rise to the ``tygers'', such as tail resonances in the energy spectrum \cite{Penati-Flach}, may be an explanation of these short-wavelength oscillations. A possible connection between ``tygers'' and our oscillations, could be the subject of a separate study.

\emph{-- Conclusions.}
In this Letter we have shown that the Fourier Energy Spectrum of the FPUT chain displays, in a wide range of specific energies, an inertial range characterised by a power-law scaling. The values of the time-dependent exponent and the time-scales involved are theoretically predicted by performing a nontrivial continuum limit of the lattice model. This procedure allows us to describe the FPUT dynamics with a pair of generalised, inviscid Burgers equations. The power-law exponent of the Fourier Energy spectrum of the chain takes the value $8/3$ at the shock time and then stabilises around $2$ before the system eventually relaxes to equipartition. These results hold for a much larger class of initial conditions than the one discussed in this Letter, as stated below \eqref{eq:uk}. In fact, the mathematical results on the asymptotics of the spectrum proven in the Supplemental Material are valid for a generic superposition of Fourier modes. Our result provides a direct relation between the FPUT dynamics and Burgers turbulence. 
Beside considerably expanding the phenomenology of the FPUT chain with an impact on the problem of relaxation to equilibrium, we believe that our results are relevant for the experimental investigations of physical systems described by the FPUT dynamics: i.e. phononic, photonic or cold atomic systems at energies higher than those at which the FPUT ``recurrence'' phenomenon has been already observed \cite{Piera}.

\bigskip

\begin{acknowledgments}
This work was partially supported by: GNFM (INdAM) the Italian national group for Mathematical Physics, the MIUR PRIN 2017 project MaQuMA cod. 2017ASFLJR and 
the MIUR-PRIN2017 project “Coarse-grained description for non-equilibrium systems and transport phenomena (CO-NEST)” No. 201798CZL.
MG and AP thank SISSA for hospitality.

\end{acknowledgments}

\nocite{*}
\providecommand{\noopsort}[1]{}\providecommand{\singleletter}[1]{#1}%



\begin{widetext}

\newpage

\newcounter{defcounter}
\setcounter{defcounter}{0}

\newenvironment{smequation}{
\addtocounter{equation}{-1}
\refstepcounter{defcounter}
\renewcommand\theequation{S-\thedefcounter}
\begin{equation}}
{\end{equation}}

\centerline{ \huge SUPPLEMENTAL MATERIAL}

\vspace{0.2cm}
\centerline{ \large to}

\vspace{0.2cm}
\centerline{ \Large ``Burgers Turbulence in the Fermi-Pasta-Ulam-Tsingou chain''}

\vspace{0.2cm}
\centerline{ \large Matteo Gallone, Matteo Marian, Antonio Ponno and Stefano Ruffo}

\section{Derivation of the continuum equations for the fields $L$ and $R$ from the lattice dynamics}

\noindent
In this section we deduce the equations of motion of the FPUT system in the continuum/thermodynamic limit $N\to\infty$, with an emphasis on their Hamiltonian form. The goal is to obtain the Hamiltonian formulation of the problem in terms of the left and right fields $L$ and $R$, that is convenient for perturbation theory.

\medskip

We first observe that, assuming the canonical variables of the form 
$q_j(t)=NQ(j/N,t/N)$ and $p_j(t)=P(j/N,t/N)$, with $Q(x,\tau)$ and $P(x,\tau)$ unknown fields of unit space period, one has 
\[
\dot{q}_j=Q_\tau(x_j,\tau)\ \ ;\ \ \dot{p}_j=P_\tau(x_j,\tau)/N\ \ ;\ \ \ x_j=j/N\ \ ;\ \ \tau=t/N\ . 
\]
Moreover, 
\[
\begin{split}
q_{j+1}-q_j=&  Q_x(x_j,\tau)+Q_{xx}(x_j,\tau)/(2N)+\cdots \\
q_j-q_{j-1}=&  Q_x(x_j,\tau)-Q_{xx}(x_j,\tau)/(2N)+\cdots \\
\end{split}
\]
Then, for any function $f$, omitting the $\tau$ dependence, one has
\[
f(q_{j+1}-q_j)-f(q_j-q_{j-1})=f'(Q_x(x_j))Q_{xx}(x_j)/N+\cdots=\frac{1}{N}[f(Q_x(x))]_x\Big|_{x=x_j}+\cdots\ ,
\]
the dots denoting everywhere terms of higher order in $1/N$. The Hamiltonian and the Hamilton equations of the FPUT chain with Hamiltonian (\ref{eq:H}) are
\begin{smequation}
\label{eq:SHeq}
\dot{q}_j=\frac{\partial H}{\partial p_j}=p_j\ \ ;\ \ \dot{p}_j=-\frac{\partial H}{\partial q_j}=
V'(q_{j+1}-q_j)-V'(q_j-q_{j-1})\ .
\end{smequation}

\noindent
In the paper we focus on the usual form of the FPUT potential $V(z)=z^2/2+\alpha z^3/3+\beta z^4/4$, as well as on the specific initial conditions
(\ref{eq:indat}). Defining the specific energy functional
\[
w[Q,P]=\lim_{N\to\infty}\frac{H}{N}=\lim_{N\to\infty}\sum_{j=1}^N\left[\frac{P^2(x_j,\tau)}{2}+V(Q_x(x_j,\tau)+\cdots)\right]\Delta x_j\ ,
\]
with $\Delta x_j=1/N$, and taking into account the expansions reported above, keeping $x_j$ finite and renaming it $x$, one finds the continuum form of the Hamilton equations in the limit $N\to\infty$
\begin{smequation}
w[Q,P]=\oint\left[\frac{P^2}{2}+V(Q_x)\right]\ dx\ ; \label{eq:w}
\end{smequation}
\begin{smequation}
Q_\tau=\frac{\delta w}{\delta P}=P\ \ ;\ \ P_\tau=-\frac{\delta w}{\delta Q}=[V'(Q_x)]_x\ . \label{eq:weqn}
\end{smequation}

\noindent
The integral $\oint$ in (\ref{eq:w}) is performed on any unit interval ($0<x_j\leq 1$) with periodic boundary conditions on the fields
$Q$ and $P$, and the functional derivatives $\delta w/\delta Q$ and $\delta w/\delta P$ are defined in the usual way \cite{Ref1}.  

In the continuum limit, the initial conditions (\ref{eq:indat}) read
\begin{smequation}
\label{eq:inQP}
Q(x,0)=\frac{a}{\pi}\cos\varphi\sin(2\pi x)\ \ ;\ \ P(x,0)=2a\sin\varphi\cos(2\pi x)\ ,
\end{smequation}

\noindent
with $a\equiv \pi A/N$ supposed to be fixed and small in the limit $A\to\infty$, $N\to\infty$. By substituting the initial condition (\ref{eq:inQP})
into (\ref{eq:w}), and taking into account that $w$ is preserved by the flow of equations (\ref{eq:weqn}) to its initial constant value $w=\varepsilon$ (as can be explicitly checked), one gets
equation (\ref{eq:epsa}), which relates the amplitude $a$ to the specific energy 
$\varepsilon$ of the system. The latter quantity, and $a$ as a consequence, are supposed to be small. 

In order to study the Hamiltonian initial value problem (\ref{eq:w}), (\ref{eq:weqn}), (\ref{eq:inQP}), we introduce the left ($L$) and right ($R$) rescaled fields 
\begin{smequation}
\label{eq:LR}
L(x,\tau)=\frac{Q_x(x,\tau)+P(x,\tau)}{a\sqrt{2}}\ \ ;\ \ 
R(x,\tau)=\frac{Q_x(x,\tau)-P(x,\tau)}{a\sqrt{2}}\ ,
\end{smequation}

\noindent
in terms of which rescaled Hamiltonian functional $h[L,R]=\lim_{N\to\infty}H/(Na^2)$ takes on the form 
\begin{smequation}
\label{eq:h}
h=\frac{\langle L^2+R^2\rangle}{2}+\frac{a\alpha\langle(L+R)^3\rangle}{6\sqrt{2}}+
\frac{a^2\beta\langle(L+R)^4\rangle}{16}\ .
\end{smequation}

\noindent
Here and henceforth $\langle f\rangle\equiv\int_0^1 f(x)dx$ denotes the spatial average of $f(x)$. The Hamilton equations 
associated to (\ref{eq:h}) are
\begin{smequation}
\label{eq:LRsys}
\left\{
\begin{split}
L_\tau=&\left(\frac{\delta h}{\delta L}\right)_x=\left[L+\frac{\alpha a}{2\sqrt{2}}(L+R)^2+
\frac{\beta a^2}{4}(L+R)^3\right]_x \\ 
R_\tau=&-\left(\frac{\delta h}{\delta R}\right)_x=-\left[R+\frac{\alpha a}{2\sqrt{2}}(L+R)^2+
\frac{\beta a^2}{4}(L+R)^3\right]_x
\end{split}\right.\ ,
\end{smequation}

\noindent
which can be obtained by direct substitution of (\ref{eq:LR}) into (\ref{eq:weqn}). In the same way, the initial condition (\ref{eq:inQP}) transform into
\begin{smequation}
\label{eq:indatLR}
\left\{
\begin{split}
L_0(x)\equiv L(x,0)=&\sqrt{2}(\cos\varphi+\sin\varphi)\cos(2\pi x) \\
R_0(x)\equiv R(x,0)=&\sqrt{2}(\cos\varphi-\sin\varphi)\cos(2\pi x)
 \end{split}
 \right.\ .
\end{smequation}

\noindent
We stress that $L_0$ has maximal amplitude $2$ for $\varphi=\pi/4$, when $R_0=0$, which defines precisely what is the left TWE. We also observe that $\oint P\ dx=\langle P\rangle$ is a constant of motion of system (\ref{eq:weqn}), and from the initial condition (\ref{eq:inQP}), it follows that $\langle P\rangle=0$, which in turn implies the conservation law $\langle Q \rangle=0$. Then, by integrating (\ref{eq:LR}), one gets $\langle L\rangle=0$ and $\langle R\rangle=0$.

\section{Perturbation theory: from equations (\ref{eq:LRsys}) to the decoupled, generalised Burgers equations (\ref{eq:larosys})}

\noindent
Let us fix a Hamiltonian functional $w[Q,P]=\oint \mathcal{W} dx=\langle \mathcal{W}\rangle$ whose density $\mathcal{W}$ is a function of the fields 
$Q_x$, $P$ and of their derivatives with respect to $x$ up to a given finite order (in the sequel, functionals are denoted by lowercase letters, whereas their densities are denoted 
by the same capital, calligraphic letter). The FPUT Hamiltonian (\ref{eq:w}) belongs to this class of functionals. Let us then consider a functional 
$f[Q,P]=\oint \mathcal{F} dx=\langle \mathcal{F}\rangle$ whose density $\mathcal{F}$ is in the same class of $\mathcal{W}$. Then, given Hamilton equations
\begin{smequation}
\label{eq:caneq}
Q_\tau=\frac{\delta w}{\delta P}\ \ ;\ \ P_\tau=-\frac{\delta w}{\delta Q}\ ,
\end{smequation}

\noindent
the time derivative of $f$ along their solutions satisfies
\begin{smequation}
\label{eq:dfdt}
\frac{df}{d\tau}=\left\langle \frac{\delta f}{\delta Q}\frac{\delta w}{\delta P}-\frac{\delta f}{\delta P}\frac{\delta w}{\delta Q}\right\rangle\equiv
\{f,w\}_c\ ,
\end{smequation}

\noindent
where we have defined the canonical Poisson bracket $\{\ ,\}_c$ relative to the canonical coordinates $Q$ and $P$. One can check that such a bracket is an actual Poisson bracket since it is bilinear, antisymmetric and satisfies Jacobi identity and Leibnitz rule. The equations of motion (\ref{eq:caneq}) read $Q_\tau=\{Q,w\}_c$, $P_\tau=\{P,w\}_c$. Everything here is completely analogous to the finite dimensional case. 

On the other hand, by applying the transformation (\ref{eq:LR}), one has $w[Q,P]\to h[L,R]=\langle \mathcal{H}\rangle$, 
$f[Q,P]\to \tilde f[L,R]=\langle\tilde{\mathcal{F}}\rangle$, where their respective densities 
$\mathcal{H}$ and $\tilde{\mathcal{F}}$ depend on $L$, $R$ and their derivatives with respect to $x$ up to a finite order. As a consequence, formula (\ref{eq:dfdt}) transforms to
\begin{smequation}
\label{eq:dftildt}
\frac{d\tilde f}{d\tau}=
\left\langle \frac{\delta \tilde f}{\delta L}\left(\frac{\delta h}{\delta R}\right)_x-
\frac{\delta \tilde f}{\delta R}\left(\frac{\delta w}{\delta L}\right)_x\right\rangle\equiv
\{\tilde f,h\}_G\ ,
\end{smequation}

\noindent
where this last equation defines the Gardner bracket $\{\ ,\}_G$. Since (\ref{eq:dfdt}) and (\ref{eq:dftildt}) express the time derivative of one and the same functional $f$ (or $\tilde f$), they imply the identity $\widetilde{\{f,w\}}_c=\{\tilde f,h\}_G$, 
where the tilde on the left means that one first computes the Poisson bracket of $f$ and $w$ with respect to $Q$ and $P$ and then performs the change of variables (\ref{eq:LR}) from $(Q,P)$ to $(L,R)$. Thus the bracket defined in (\ref{eq:dftildt}) is also a Poisson bracket, being just the transformed of the canonical Poisson bracket defined in (\ref{eq:dfdt}). The equations of motion
associated to $h$ in the latter structure read $L_\tau=\{L,h\}_G$, $R_\tau=\{R,h\}_G$, that for the FPUT Hamiltonian (\ref{eq:h}) yields just the equations (\ref{eq:LRsys}).

Equation (\ref{eq:dftildt}) can be symbolically solved by defining the operator $\mathcal{L}_h=\{\ ,h\}_G$ such that
$\mathcal{L}_h\tilde f=\{\tilde f,h\}_G$, namely
\begin{smequation}
\label{eq:ftiltau}
\tilde f(\tau)=e^{\mathcal{L}_h\tau}\tilde f(0)=\left(1+\tau\mathcal{L}_h+\frac{\tau^2}{2}\mathcal{L}_h^2+\cdots\right)
\tilde f(0)\ ,
\end{smequation}

\noindent
where the exponential of $\tau\mathcal{L}_h$ is defined by its series, the dots denoting higher order terms. The fundamental result used below is the following: the flow $e^{\tau\mathcal{L}_h}$ of $h$, with $\mathcal{L}_h=
\{\ ,h\}_G$, preserves the Poisson bracket $\{\ ,\}_G$ itself, in the sense that
$e^{\tau\mathcal{L}_h}\{a,b\}_G=\left\{e^{\tau\mathcal{L}_h}a,e^{\tau\mathcal{L}_h}b\right\}_G$
for any pair of functionals $a=\langle \mathcal{A}\rangle$ and $b=\langle \mathcal{B}\rangle$ in the given class
\cite{Ref2}.
As a final remark, notice that in the above treatment no special form of $h$ (and the other functionals) was considered, so that the flow of \emph{any} functional $h$ preserves the Hamiltonian structure defining it (this is completely general, the Gardner structure being just an example). 

We now make use of the above tools to transform the Hamiltonian (\ref{eq:h}) of the FPUT system and decouple the corresponding Hamilton equations to second order in the small parameter $a$. The Hamiltonian (\ref{eq:h}) can be obviously ordered as
\begin{smequation}
h=h_0+ah_1+a^2h_2
\end{smequation}
where 
\begin{smequation}
\label{eq:h012}
h_0=\frac{\langle L^2+R^2\rangle}{2}\ \ ;\ \ h_1=\frac{\alpha\langle(L+R)^3\rangle}{6\sqrt{2}}\ \ ;\ \ 
h_2=\frac{\beta\langle(L+R)^4\rangle}{16}\ .
\end{smequation}

\noindent
The equations of motion of $h_0$ have the form $L_\tau=L_x$, $R_\tau=-R_x$, whose solution is
$L(x,\tau)=L_0(x+\tau)$, $R(x,\tau)=R_0(x-\tau)$, i.e. the left and right translation of the initial condition, respectively.
Notice that the latter solution is periodic in time, with period one, for any space periodic initial condition $(L_0,R_0)$ with period one. Then, with the notation introduced above, the flow of $h_0$ at time $s$ is defined by
\begin{smequation}
\label{eq:flow0}
(L_0(x+s),R_0(x-s))=e^{s\mathcal{L}_0}(L_0(x),R_0(x))\ \ ;\ \ \mathcal{L}_0=\{\ ,h_0\}_G\ .
\end{smequation}

\noindent
Notice that, since the flow of $h_0$ has period one, $e^{\mathcal{L}_0}=1$. 

We now build up a transformation of the fields $\mathcal{C}_a:(L,R)\mapsto(\lambda,\rho)=\mathcal{C}_a(L,R)$, smoothly dependent on $a$ and close to the identity ($\mathcal{C}_0(L,R)=(L,R)$), by composing two Hamiltonian flows, corresponding to two unknown generating Hamiltonians, $g_1$ and $g_2$, as follows.  
By defining $\mathcal{L}_1=\{\ ,g_1\}_G$, $\mathcal{L}_2=\{\ ,g_2\}_G$, we set
\begin{smequation}
\label{eq:Cacomp}
(L,R)=\mathcal{C}^{-1}_a(\lambda,\rho)=e^{a^2\mathcal{L}_2}e^{a\mathcal{L}_1}(\lambda,\rho)\ .
\end{smequation}

\noindent
The following conditions uniquely determine $g_1$ and $g_2$ \cite{hptpde}.
\begin{enumerate}
\item The transformed Hamiltonian $\tilde h=h\circ \mathcal{C}_a^{-1}=e^{a^2\mathcal{L}_2}e^{a\mathcal{L}_1}h$
is in normal form with respect to $h_0$ to second order in $a$, namely
$\tilde h=h_0+a \tilde{h}_1 + a^2\tilde{h}_2+O(a^3)$, with $\{\tilde{h}_1,h_0\}_G=\{\tilde{h}_2,h_0\}_G=0$ (i.e.
$\tilde{h}_1$ and $\tilde{h}_2$ are first integrals of $h_0$).
\item $g_1$ and $g_2$ have zero average on the unperturbed flow of $h_0$: 
$\int_0^1e^{s\mathcal{L}_0}g_1 ds=\int_0^1e^{s\mathcal{L}_0}g_2 ds=0$.
\end{enumerate}
For the transformed Hamiltonian, expanding the exponentials, one gets
\begin{eqnarray}
\tilde h&=&e^{a^2\mathcal{L}_2}e^{a\mathcal{L}_1}(h_0+a h_1+a^2 h_2)=\nonumber\\
&=& h_0+a(\mathcal{L}_1h_0+h_1)+
a^2\left(\mathcal{L}_2h_0+\mathcal{L}_1h_1+\frac{1}{2}\mathcal{L}_1^2h_0+h_2\right)+O(a^3)=\nonumber\\
&=&h_0+a \tilde{h}_1 + a^2\tilde{h}_2+O(a^3)\ .\nonumber
\end{eqnarray}
Thus, taking into account that $\mathcal{L}_1h_0=-\mathcal{L}_0g_1$ and $\mathcal{L}_2h_0=
-\mathcal{L}_0g_2$, one finds the two \emph{homological equations}
\begin{smequation}
\label{eq:homo12}
\begin{split}
\mathcal{L}_0g_1= & h_1-\tilde{h}_1\ ; \\
\mathcal{L}_0g_2= & \mathcal{L}_1h_1+\frac{1}{2}\mathcal{L}_1^2h_0+h_2-\tilde{h}_2\ ,
\end{split}
\end{smequation}

\noindent
for the four unknowns $g_1$, $g_2$, $\tilde{h}_1$ and $\tilde{h}_2$. The solution to equations (\ref{eq:homo12}) can be obtained taking into account the following technical points. First,
\begin{smequation}
\label{eq:step1}
\int_0^1e^{s\mathcal{L}_0}\mathcal{L}_0g_i\ ds=\int_0^1\frac{d}{ds}e^{s\mathcal{L}_0}g_i\ ds=
\left(e^{\mathcal{L}_0}-1\right)g_i=0\ ,\ \ (i=1,2)
\end{smequation}

\noindent
since the unperturbed flow $e^{s\mathcal{L}_0}$ has period one. Second, 
\begin{smequation}
\label{eq:step2}
e^{s\mathcal{L}_0}\tilde{h}_i=\left(1+s\mathcal{L}_0+\frac{s^2}{2}\mathcal{L}_0^2+\cdots\right)\tilde{h}_i= 
\tilde{h}_i\ ,\ \ (i=1,2)
\end{smequation}
since $\mathcal{L}_0\tilde{h}_i=\{\tilde{h}_i,h_0\}_G=0$, which is required by the definition of normal form.
Third, 
\begin{smequation}
\label{eq:step3}
\int_0^1se^{s\mathcal{L}_0}\mathcal{L}_0g_i\ ds=se^{s\mathcal{L}_0}g_i\Big|_{0}^1
-\int_0^1e^{s\mathcal{L}_0}g_i\ ds=g_i\ ,\ \ (i=1,2)
\end{smequation}

\noindent
since we posed the condition that the average of the generating Hamiltonians $g_i$ on the unperturbed flow vanishes (this is a choice: the normal form is not unique). By taking into account the steps (\ref{eq:step1}), (\ref{eq:step2}) and (\ref{eq:step3}), one obtains the solution of the first of equations
(\ref{eq:homo12}), namely
\begin{smequation}
\label{eq:htil1g1}
\tilde{h}_1=\int_0^1e^{s\mathcal{L}_0}h_1\ ds\ \ ;\ \ 
g_1=\int_0^1 se^{s\mathcal{L}_0}(h_1-\tilde{h}_1)\ ds\ .
\end{smequation}

Before solving the second of equations (\ref{eq:homo12}) in an analogous way, it is convenient to substitute 
$\mathcal{L}_1h_0=\tilde{h}_1-h_1$ into its right hand side, to get
$\mathcal{L}_0g_2= \frac{1}{2} \mathcal{L}_1h_1+h_2-\tilde{h}_2+\frac{1}{2}\mathcal{L}_1\tilde{h}_1$. Now,
the average of $\mathcal{L}_1\tilde{h}_1$ vanishes:
\[
\int_0^1 e^{s\mathcal{L}_0}\mathcal{L}_1\tilde{h}_1\ ds=\int_0^1 e^{s\mathcal{L}_0}\{\tilde{h}_1,g_1\}_G\ ds=
\left\{\tilde{h}_1,\int_0^1e^{s\mathcal{L}_0}g_1ds\right\}_G=0\ ,
\]
by (\ref{eq:step2}) and the bilinearity of the Poisson bracket. Thus, the average of the second of equations
(\ref{eq:homo12}) yields
\begin{smequation}
\label{eq:htil2}
\tilde{h}_2=\int_0^1e^{s\mathcal{L}_0}\left(h_2+\frac{1}{2}\mathcal{L}_1h_1\right)ds=
\int_0^1e^{s\mathcal{L}_0}\left(h_2+\frac{1}{2}\{h_1,g_1\}_G\right)ds\ .
\end{smequation}

We do not report here the expression of the generating Hamiltonian $g_2$, since it is not used neither for the computation of the Hamiltonian to second order, nor it is useful for the later transformation of initial data.

One has now to explicitly compute the quantities (\ref{eq:htil1g1}) and (\ref{eq:htil2}), where the 
functions $h_1$ and $h_2$ given in (\ref{eq:h012}), have to be expressed in the new field variables $\lambda(x,\tau)$ and $\rho(x,\tau)$. By its definition (\ref{eq:flow0}), the action of 
$e^{s\mathcal{L}_0}$ on a monomial in $\lambda$
and $\rho$ is simple:
\[
e^{s\mathcal{L}_0}\lambda^m(x,\tau)\rho^n(x,\tau)=\lambda^m(x+s,\tau)\rho^n(x-s,\tau)
\] 
In order to explicitly compute $\tilde{h}_1$, $g_1$ and $\tilde{h}_2$ one needs the following relations, valid for any pair of functions $F$ and $G$ of space period one:
\[
\int_0^1\int_0^1F(x+s)G(x-s)\ dx\ ds=\int_0^1F(x)\ dx\ \int_0^1 G(x)\ dx\ ;
\]
\begin{smequation}
\int_0^1\int_0^1s\ F(x+s)G(x-s)\ dx\ ds=\frac{1}{2}\left(\int_0^1F(x)\ dx\ \int_0^1 G(x)\ dx+
\int_0^1G(x)\partial_x^{-1} F(x)\ dx\right)\ ,
\end{smequation}

\noindent
which can be proved by expressing $F$ and $G$ in Fourier series. The antiderivative $\partial_x^{-1}$ appearing above is defined as: $\partial_x^{-1}F(x)=\sum_{k\neq0}\hat F_k/(\imath2\pi k)e^{\imath2\pi kx}$. The antiderivative is skew-symmetric under integration.

After elementary, though a bit long computations (that involve also the explicit determination of $g_1$; see (\ref{eq:g1}) below), one finds the explicit expression of the normal form Hamiltonian 
$\tilde{h}=h_0+a\tilde{h}_1 +a^2 \tilde{h}_2+O(a^3)$, namely
\begin{smequation}
\label{eq:htilrep}
\begin{split}
\tilde h=&\frac{\langle\lambda^2+\rho^2\rangle}{2}+a\left(\alpha \frac{\langle\lambda^3+
\rho^3\rangle}{6\sqrt{2}}\right) + \\
& +a^2\left[\left(2\beta-\alpha^2\right)\frac{\langle\lambda^4+\rho^4\rangle}{32}
+\alpha^2\frac{\langle\lambda^2\rangle^2+\langle\rho^2\rangle^2}{32} 
+\left(3\beta-2\alpha^2\right)\frac{\langle\lambda^2\rangle\langle\rho^2\rangle}{8}\right]+O(a^3)\ .
\end{split}
\end{smequation}
\noindent
The equations of motion associated to this Hamiltonian, up to terms $O(a^3)$, are
\begin{smequation}
\label{eq:larhoev}
\left\{
\begin{split}
& \lambda_\tau=\left(\frac{\delta \tilde h}{\delta \lambda}\right)_x= 
\left[c_l+\frac{a\alpha}{\sqrt{2}}\lambda+
\frac{3a^2\alpha^2}{4}\left(\frac{\beta}{\alpha^2}-\frac{1}{2}\right)\lambda^2\right]\lambda_x \\
& \rho_\tau=-\left(\frac{\delta \tilde h}{\delta \rho}\right)_x=
-\left[c_r+\frac{a\alpha}{\sqrt{2}}\rho+
\frac{3a^2\alpha^2}{4}\left(\frac{\beta}{\alpha^2}-\frac{1}{2}\right)\rho^2\right]\rho_x
\end{split}
\right.\ ,
\end{smequation} 

\noindent
where the left and right translation velocities $c_l$ and $c_r$ are given by
\begin{smequation}
\begin{split}
& c_l=1+\frac{\alpha^2a^2}{8}\langle\lambda^2\rangle+
\frac{3a^2}{4}\left(\beta-\frac{2\alpha^2}{3}\right)\langle\rho^2 \rangle\ ; \\
& c_r=1+\frac{\alpha^2a^2}{8}\langle\rho^2\rangle+
\frac{3a^2}{4}\left(\beta-\frac{2\alpha^2}{3}\right)\langle\lambda^2 \rangle\ .
\end{split}
\end{smequation}

\noindent
Notice that $\langle \lambda^2\rangle$ and $\langle \rho^2\rangle$ are first integrals of system
(\ref{eq:larhoev}), so that the above velocities are constant. Moreover, the field transformation defined by
$\lambda(x,\tau)=\lambda'(x+c_l\tau,\tau)$, $\rho(x,\tau)=\rho'(x-c_r\tau,\tau)$, removes the translation terms 
$c_l\lambda_x$ and $-c_r\rho_x$ in system (\ref{eq:larhoev}), which completely decouples the two equations.
We thus set $c_l=c_r=0$ without any loss of generality (which also amounts to erase all the terms containing 
$\langle\lambda^2\rangle$, $\langle\rho^2\rangle$ and their powers in the Hamiltonian (\ref{eq:htilrep})).  
We have thus justified the form of the equations (\ref{eq:larosys}).

Concerning the initial conditions $(\lambda_0,\rho_0)$ satisfied by the fields $\lambda$ and $\rho$, one deduces them as follows.
The first generating Hamiltonian $g_1$, defining the canonical transformation to first order in $a$, and necessary
to compute $\tilde{h}_2$, is 
\begin{smequation}
\label{eq:g1}
g_1=\frac{\alpha}{4\sqrt{2}}\ \left\langle\rho\partial_x^{-1}(\lambda^2)+\rho^2\partial_x^{-1}\lambda\right\rangle\ .
\end{smequation} 

\noindent
One can then express the new fields $\lambda$ and $\rho$ in terms of the old ones, $L$ and $R$, by inverting 
(\ref{eq:Cacomp}), namely 
\[
\lambda= e^{-a\mathcal{L}_1}L=L-a\{L,g_1\}+O(a^2)\ \ ;\ \ 
\rho= e^{-a\mathcal{L}_1}R=R-a\{R,g_1\}+O(a^2)\ .
\]
The final result is
\begin{smequation}
\label{eq:Caoa}
\begin{split}
\lambda=&  L+\frac{\alpha a}{4\sqrt{2}}\left(R^2-\langle R^2\rangle\right)+
\frac{\alpha a}{2\sqrt{2}}(LR+L_x\partial_x^{-1}R)+O(a^2)\ ;\\
\rho=& R+\frac{\alpha a}{4\sqrt{2}}\left(L^2-\langle L^2\rangle\right)+
\frac{\alpha a}{2\sqrt{2}}(LR+R_x\partial_x^{-1}L)+O(a^2)\ .
\end{split}
\end{smequation}

\noindent 
Substituting in the latter expression the initial condition (\ref{eq:indatLR}), with $\theta=\varphi-\pi/4$, and neglecting the remainder $O(a^2)$, one gets
\begin{smequation}
\label{eq:indatlarho}
\left\{
\begin{split}
\lambda_0= & 2\cos\theta\cos(2\pi x)+\frac{a\alpha(\sin^2\theta-2\sin2\theta)}{2\sqrt{2}}\cos(4\pi x)\\
\rho_0 = &-2\sin\theta\cos(2\pi x)+\frac{a\alpha(\cos^2\theta-2\sin2\theta)}{2\sqrt{2}}\cos(4\pi x)
\end{split}
\right.\ ,
\end{smequation}

\noindent
We finally observe that the transformation (\ref{eq:Caoa}) preserves the space average of the fields, so that $\langle \lambda\rangle=\langle L\rangle=0$ and $\langle \rho\rangle=\langle R\rangle=0$.

\section{Shock time computation: derivations of formulas (\ref{eq:ts}), (\ref{eq:Fmu}) and (\ref{eq:mu})}

\noindent
Our transformed system (\ref{eq:larosys}), (\ref{eq:indatlarho}), has the form of two decoupled, generalised Burgers equations, with a given initial datum. Now, given the generalised Burgers equation $u_\tau=f(u)u_x$, with initial datum $u_0(x)$, its solution $u(x,\tau)$ is implicitly defined by the equation $u-u_0(x+f(u)\tau)=0$ (which can be checked by direct inspection). The latter identity admits an explicit solution if the implicit function theorem applies, namely - taking the derivative with respect to $u$ - if
\[
1-u_0'(x+f(u)\tau)f'(u)\tau=1-\tau\frac{d}{d\xi}f(u_0(\xi))\neq0\ ;\ \ \xi\equiv x+f(u)\tau\ .
\]
The above condition is satisfied for all $\tau$ in the interval $[0,\tau_s[$, where $\tau_s$, the shock time, is given by 
\begin{smequation}
\label{eq:16}
\frac{1}{\tau_s}= \max_x\frac{d}{dx}f(u_0(x))\ .
\end{smequation}

\noindent
Now, \eqref{eq:laroeq} consists of two independent equations, and the shock time of the FPUT system is given by 
$\tau_s=\min\{\tau_s^l,\tau_s^r\}$, whereas the left and right shock times $\tau^l_s$ and $\tau_s^r$ are given by
\begin{smequation}
\label{eq:taulr}
\frac{1}{\tau_s^l}=\max_{x\in[0,1]}\left[\frac{d}{dx}\Phi(\lambda_0(x))\right]\ \ ;\ \ 
\frac{1}{\tau_r^l}=\max_{x\in[0,1]}\left[-\frac{d}{dx}\Phi(\rho_0(x))\right]\ .
\end{smequation}

\noindent
Here $\Phi$ is the function defined in (\ref{eq:Phi}), $\lambda_0$ and $\rho_0$ are given in (\ref{eq:indatlarho})
and  in $\Phi(\lambda_0(x))$ and $\Phi(\rho_0(x))$ one has to consistently neglect terms $O(a^3)$. The explicit computation of of $\tau_s^l$ and $\tau_s^r$ in \eqref{eq:taulr} for the left shock time yields
\begin{smequation}
\label{eq:taul}
\tau_s^l=\left(\frac{1}{2\pi\sqrt{2}a\alpha}\right)\frac{1}{\cos\theta}
\sqrt{\frac{32\mu^2}{\sqrt{1+32\mu^2}-1+16\mu^2}}\frac{4}{\sqrt{1+32\mu^2}+3}\ ,
\end{smequation}

\noindent
where 
\[
\mu=\frac{a\alpha}{2\sqrt{2}}\cos\theta\left[
\tan^2\theta-4\tan\theta+6\left(\frac{\beta}{\alpha^2}-\frac{1}{2}\right)\right]\ , 
\]
whereas the right shock time is given by
\begin{smequation}
\label{eq:taur}
\tau_s^r=\left(\frac{1}{2\pi\sqrt{2}a\alpha}\right)\frac{1}{|\sin\theta|}
\sqrt{\frac{32\eta^2}{\sqrt{1+32\eta^2}-1+16\eta^2}}\frac{4}{\sqrt{1+32\eta^2}+3}\ ,
\end{smequation}

\noindent
where
\[
\eta=\frac{a\alpha}{2\sqrt{2}}\sin\theta\left[
\cot^2\theta-4\cot\theta+6\left(\frac{\beta}{\alpha^2}-\frac{1}{2}\right)\right]\ . 
\]

\noindent
Now, in the range $-\pi/4\leq\theta\leq\pi/4$, for $a$ small enough, and any $\alpha$, $\beta$, 
the inequality $\tau_s^l\leq\tau_s^r$ holds,
the equality being valid only for $\theta=\pm\pi/4$.  It follows that in the same range of $\theta$ and $a$ and any
$\alpha$, $\beta$, $\tau_s=\min\{\tau_s^l,\tau_s^r\}=\tau_s^l$. Recalling that $\tau=t/N$, one gets formulas
(\ref{eq:ts}), (\ref{eq:Fmu}) and (\ref{eq:mu}).

\section{FES asymptotics: derivation of formulas (\ref{eq:uk}) and (\ref{eq:fesgen2})}

\noindent
First of all, given the generalised Burgers equation $u_\tau=f(u)u_x$, we prove that expressing its solution 
in Fourier series, namely $u(x,\tau)=\sum_k\hat{u}_k(\tau)e^{\imath 2\pi kx}$, the Fourier coefficient $\hat u_k(\tau)$
can be expressed in terms of the initial condition $u_0(x)$ by the explicit formula (\ref{eq:uk}).
Taking into account that $u=u_0(x+\tau f(u))$, and introducing the variable $\xi$ such that $\xi=x+\tau f(u_0(\xi))$, one has
\begin{smequation}
\begin{split}
\hat{u}_k(\tau)=& \oint e^{-\imath2\pi kx}u_0(x+f(u)\tau)\ dx = 
  \oint e^{-\imath2\pi k[\xi-\tau f(u_0(\xi))]} u_0(\xi) \frac{d}{d\xi}\left[\xi-\tau f(u_0(\xi))\right]\ d\xi= \\
=& \frac{1}{-\imath2\pi k}\oint u_0(\xi)\frac{d}{d\xi}\left[e^{-\imath2\pi k[\xi-\tau f(u_0(\xi))]}\right]\ d\xi =
 \frac{1}{\imath2\pi k}\oint u'_0(\xi)e^{-\imath2\pi k[\xi-\tau f(u_0(\xi))]}\ d\xi\ .
\end{split}
\end{smequation}
Applying this to the first of equations (\ref{eq:larosys}), with $f=\Phi$, yields (\ref{eq:uk}).

\medskip

We now prove the following theorem: \textit{Let the initial condition $u_0(x)$ of the generalised Burgers equation
$u_\tau=f(u)u_x$ satisfy the following three conditions: 
\begin{enumerate}
\item[(i)] $u_0(x)=\sum_{n=-M}^Mc_ne^{\imath2\pi nx}$; $c_0=0$, $M$ finite;
\item[(ii)] $df(u_0(x))/dx$ admits a finite number $m$ of absolute maximum points $x_1,\dots,x_m\in[0,1[$;
\item[(iii)] $\gamma_j\equiv d^3f(u_0(x_j))/dx^3\neq0$.
\end{enumerate}
Then, the asymptotic formula 
\begin{smequation}
\label{eq:fesgen}
|\hat u_k(\tau_s)|^2\sim\left|\sum_{j=1}^m\frac{u_0'(x_j)e^{-\imath2\pi k[x_j-\tau_sf(u_0(x_j))]}}{
(9\pi \tau_s \gamma_j)^{1/3}\Gamma(2/3)}\right|^2k^{-8/3}
\end{smequation}
holds as $k\to+\infty$, where $\Gamma(2/3)$ is the Euler gamma function at $2/3$.}

We start  by the expression (\ref{eq:uk}) for $\hat u_k(\tau)$, and split the unit integration interval into $m$ disjoint subintervals $I_1,\dots,I_m$, such that $I_j$ contains only the maximum point $x_j$ in its interior. Thus
\begin{smequation}
\label{eq:uktau}
u_k(\tau)=\frac{1}{k}\sum_{n=-M}^Mc_nn\sum_{j=1}^m\int_{I_j}e^{\imath2\pi nx}
e^{-\imath2\pi k[x-\tau f(u_0(x))]}\ dx\ .
\end{smequation}

\noindent
In the asymptotics $k\to\infty$ each of the integrals on $I_j$ is treated with the method of stationary phase \cite{Ref3}. One has to take into account that, by the definition (\ref{eq:16}) of the shock time $\tau_s$, and by the hypotheses (ii) and (iii) above, in the interval $I_j$
\[
x-\tau_s f(u_0(x))=x_j-\tau_s f(u_0(x_j))-\frac{\tau_s\gamma_j}{6}(x-x_j)^3+O((x-x_j)^4)\ .
\]
Thus, if $I_j=[x_j-a_j,x_j+b_j[$, changing variable to $u=x-x_j$, for $k\gg 1$ and $\tau=\tau_s$ one finds
\begin{smequation}
\label{eq:Ij}
\begin{split}
& \int_{I_j}e^{\imath2\pi nx}e^{-\imath2\pi k[x-\tau_s f(u_0(x))]}\ dx=  e^{\imath2\pi nx_j}
e^{-\imath2\pi k[x_j-\tau_s f(u_0(x_j))]}\int_{-a_j}^{b_j}e^{\imath2\pi n u}
e^{\imath\frac{\pi \tau_s\gamma_j}{3}ku^3+O(ku^4)}\ du=\\
&=\  \frac{e^{\imath2\pi nx_j}e^{-\imath2\pi k[x_j-\tau_s f(u_0(x_j))]}}{(\pi|\gamma_j|\tau_s k)^{1/3}}
\int_{-a_j(\pi|\gamma_j|\tau_sk)^{1/3}}^{b_j(\pi|\gamma_j|\tau_s k)^{1/3}}
e^{\imath \frac{2\pi n}{(\pi|\gamma_j|\tau_s k)^{1/3}}z}
e^{\imath \mathrm{sgn}(\gamma_j)\frac{z^3}{3}+O(z^4/k^{1/3})}\ dz\sim \\
&\sim \  \frac{e^{\imath2\pi nx_j}e^{-\imath2\pi k[x_j-\tau_s f(u_0(x_j))]}}{(\pi\gamma_j\tau_s k)^{1/3}}
\int_{-\infty}^{+\infty} e^{\imath\mathrm{sgn}(\gamma_j)\frac{z^3}{3}}\ dz=
\frac{e^{\imath2\pi nx_j}e^{-\imath2\pi k[x_j-\tau_s f(u_0(x_j))]}}{(\pi\gamma_j\tau_s k)^{1/3}}\ 
\frac{2\pi}{3^{2/3}\Gamma(2/3)}\ .
\end{split}
\end{smequation}

\noindent
The change of variable $z=(\pi|\gamma_j|\tau_sk)^{1/3} z$ has been done in passing from the second to the third line, and $\mathrm{sgn}(\gamma_j)$ is the sign function. The last step is obtained by observing that
$\int_{-\infty}^\infty\cos(z^3/3)dz=2\pi Ai(0)$, where $Ai(0)=3^{-2/3}/\Gamma(2/3)$ is the value of the Airy function at zero. Inserting (\ref{eq:Ij}) into (\ref{eq:uktau}) at $\tau=\tau_s$ one gets
\[
\hat u_k(\tau_s)\sim -\imath\left[\sum_{j=1}^m
\frac{u_0'(x_j)e^{-\imath2\pi k[x_j-\tau_s f(u_0(x_j))]}}{(9\pi|\gamma_j|\tau_s)^{1/3}\Gamma(2/3)}\right] k^{-4/3}\ ,
\] 
whose square modulus yields (\ref{eq:fesgen}).

Finally, formula (\ref{eq:fesgen2}) for the normalised FES is obtained by proving that the function
$d\Phi(\lambda_0(x))/dx$, which enters the definition (\ref{eq:taulr}) of the shock time
(\ref{eq:taul}), displays a single, absolute maximum point $\hat x$ if $a$ is small enough. Then, formula (\ref{eq:fesgen})
simplifies to
\begin{smequation}
\label{eq:laktaus}
|\hat \lambda_k(\tau_s)|^2\sim\left|\frac{u_0'(\hat x)}{
(9\pi \tau_s d^3\Phi(\lambda_0(\hat x))/dx^3)^{1/3}\Gamma(2/3)}\right|^2k^{-8/3}= C\ k^{-8/3}\ ,
\end{smequation}

\noindent
where the constant $C$ is independent of $k$. Assuming that the form of the FES at the shock time be given by 
$C k^{-8/3}$ for all $k\geq 1$, the normalised FES is given by
\[
\frac{E_k(t_s)}{\sum_{k>0}E_k(t_s)}=\frac{|\hat \lambda_k(\tau_s)|^2}{\sum_{k>0}|\hat \lambda_k(\tau_s)|^2}=
\frac{k^{-8/3}}{\sum_{k>0}k^{-8/3}}=\frac{k^{-8/3}}{\zeta_R(8/3)}\ ,
\] 
where $\zeta_R(s)=\sum_{k>0}k^{-s}$ is the Riemann zeta function. One finds numerically that
$\zeta_R(8/3)=1.28419\dots$, whose reciprocal is $0.77870\dots$, which justifies formula (\ref{eq:fesgen2}). For the extreme values of $\theta=\pm\pi/4$, where both the left and the right channel contribute to the spectrum, the formula is the same because the two contributions to the FES are identical.

%
%
%
%
%

%
%
%

\newpage

\end{widetext}

\begin{thebibliography}{10}
\bibitem{cinfinity} For $f \in C^\infty[0,1]$, $f$ periodic, all the derivatives are finite. Since $d^nf(x)/dx^n = 
\sum_{k} (\imath 2 \pi  k)^n \hat{f}_ke^{\imath 2 \pi k x}$, one also has 
$\hat{f}_k = \frac{1}{(\imath 2 \pi  k)^n} \int_0^1 e^{-\imath 2 \pi k x}(d^nf(x)/dx^n)\, dx$ implying that 
$\hat{f}_k$ decays faster than any power in $k$.
\end{thebibliography}
\end{document}